\documentclass[twocolumn]{article}
\usepackage{color}

\author{
  Markus Kloimwieder \\ 
  Unaffiliated \\
  Vienna, 1010, Austria
  \and
  Christoph Gadermaier \\ 
  Dipartimento di Fisica, Politecnico di Milano \\ 
  Milano, 20133, Italy
}
\title{A functional scripting interface to an object oriented C++ library}

\date{December 2023}

\begin{document}
\twocolumn[
  \begin{@twocolumnfalse}
    \maketitle
    \begin{abstract}
      The object oriented programming paradigm is widely used in science and engineering. Many open and commercial libraries are written in C++, and increasingly provide bindings to Python, which is much easier to learn, but still partly encourages the use of object oriented programming. However, scientific ideas are much more directly and meaningfully expressed in the purely functional programming paradigm. Here, we take a best practice example – CERN’s Python binding for its ROOT library, designed to handle the enormous amounts of data generated by the world’s largest particle accelerator – and translate a simple segment of its tutorial into Clojure, a functional language from the Lisp family. The code examples demonstrate how a purely functional language straightforwardly expresses scientific ideas. Subsequently, we develop a compiled Lisp-C++ interoperation layer to access the ROOT library exclusively via functional code. To preserve the expressivity of the Lisp code, the type hints necessary for C++ code generation are stored in a separate file. The interop system presented here is a generic framework that, when provided with a suitable file of type hints, facilitates access to methods of arbitrary C++ libraries and platforms like real-time microcontrollers.
      \newline
    \end{abstract}
    \end{@twocolumnfalse}]

\section{INTRODUCTION}
Large amounts of code for applications in science and engineering are written using the object-oriented (OO) programming paradigm. A best practice example is the ROOT library developed by CERN.$^{\cite{brun1997root}}$ It was originally designed for particle physics data analysis, but it is also used in other applications such as astronomy and data mining. ROOT is written in C++ and includes an interpreter called Cling that allows the users to run their own C++ scripts. Yet, since C++ is a complicated language with a steep learning curve, CERN also provides ROOT bindings for Python, which is a very popular multi-purpose language and is much easier to learn. However, Python has been originally designed to replace shell scripts, used for day-to-day system administration. Python has not been designed to express scientific ideas.

A very suitable way to express scientific ideas in computer programs is provided by the functional programming paradigm, as demonstrated in the books “Structure and Interpretation of Classical Mechanics”$^{\cite{sussman2015structure}}$ and “Functional Differential Geometry”$^{\cite{sussman2013functional}}$ as well as in the paper “The role of programming in the formulation of ideas”$^{\cite{sussman2002role}}$. These seminal works are coded in MIT Scheme, a dialect of Lisp. Unfortunately, MIT Scheme does not have a large ecosystem and is also hard to install on MS-Windows systems. A Lisp dialect with a larger ecosystem is Clojure.$^{\cite{hickey2008clojure}}$ It runs on the Java virtual machine (JVM), which is a household name in the IT industry. However, the JVM has not gained traction in the natural sciences; the reason might be that the JVM is a very memory consuming runtime environment that is not well suited for running small scripts.

A recent ongoing attempt to circumvent the JVM is jank,$^{\cite{wilkerson2022jank}}$ which aims at implementing the Clojure dialect on ROOT’s C++ interpreter Cling. It has the potential to open up the uniformity, simplicity and expressiveness of functional programming to ROOT users. To this aim, it is crucial to implement an idiomatic Lisp-C++ interoperation which compiles to native binary code, effectively resulting in a purely functional C++ layer on top of ROOT. The major difficulty here is this: Lisp is dynamically typed whereas C++ is statically typed.

In the respective Lisp-Java interoperation of JVM-Clojure, this difficulty is overcome via so-called type hints directly written into the users’ Clojure code, obstructing the expressive syntax. For our Lisp-C++ interop we also use type hints but contrary to the current Java approach, those hints are not visible in the syntax of the user code, which thus maintains its full expressiveness. We avoid obstructing our Lisp code by storing the type hints in a separate data file. Those are then checked at compile-time as well as at runtime within the small executable binary.

The article is structured as follows: first we take a publicly available Python tutorial and translate it to Lisp. The first part of this nine-line script does not yet call into ROOT; its translation is merely meant to demonstrate the principles underlying the functional paradigm. The second part of the tutorial indeed involves ROOT and here we explain our Lisp-C++ interoperation method. After a short recap, in the third part of the article we describe our unique approach for calling into different ROOT subroutines by the same function name, thus addressing interoperation with C++ “runtime polymorphism”. Finally, we discuss the scope and limitations of the presented interop as a generic tool intended to access arbitrary C++ libraries, which are widely used in science, engineering, and quantitative finance.

\section{THE TUTORIAL}
This article frequently refers to a specific publicly available tutorial script,$^{\cite{root2023python}}$ which CERN provides as an introduction to the ROOT library. The example reproduced here shows how to plot a simple graph – a straight line – with ROOT in nine lines of Python:
{\color{magenta}\begin{verbatim}
import ROOT

class Linear:
    def __call__(self, arr, par):
        return par[0] + arr[0]*par[1]

# create a linear function
l = Linear()
f = ROOT.TF1('pyf2', l, -1., 1., 2)
f.SetParameters(5., 2.)

# plot the function
c = ROOT.TCanvas()
f.Draw()
\end{verbatim}}
An integral part of our discussion is the translation and comparison of this Python code to Clojure, a dialect of the Lisp family of languages currently used mainly for commercial web applications.$^{\cite{clojure2023state}}$ Despite both being multi-paradigm languages, Python and Clojure nevertheless represent two different paradigms in software design: “OO” and “functional”, respectively. The purpose of our translation is to demonstrate and encourage the usage of functional programming in the natural sciences and beyond.

\section{THE HEADER}
\subsection{Python interop via cppyy}
While the CERN tutorial to plot the graph of a straight line is written in Python, the ROOT library is programmed in C++, the major systems programming language known for its compact executable binaries. The necessary interoperation between the scripting language and the library is done via cppyy, whose runtime "automatic binding provision requires no direct support from library authors".$^{\cite{kundu2023efficient}}$ In order to make cppyy available within a users’ script, its main module needs to be imported, which accordingly is done in the first statement of the tutorial code:
{\color{magenta}\begin{verbatim}
import ROOT
\end{verbatim}}
cppyy operates behind the scenes as subsequent Python code is interpreted and executed. This operation of cppyy at runtime is a very flexible way to solve all problems stemming from the fundamental difference between Python and C++: Python is dynamically typed whereas C++ is statically typed. With cppyy, the types of the Python variables are determined at runtime and C++ code is generated on the fly to be then processed by ROOT’s C++ interpreter Cling.

\subsection{Translation to LISP}
With Lisp, as opposed to Python, we do not use a runtime bridge or Cling to access ROOT, which means that within this article we nowhere make use of the advanced "automatic binding" feature of cppyy. Instead, Lisp code is compiled (or converted) to C++ code ahead of its execution time. Since Lisp is also a dynamically typed language, we will build a prototype for a mechanism to dispatch into pre-compiled C++ code according to runtime types, such a mechanism usually being termed “runtime polymorphism”.

For compilation we use Ferret, a free Lisp implementation designed to be used in real time embedded control systems, supporting a subset of Clojure  to generate executable binaries for PC as well as microcontrollers with “as little as 2KB of RAM”.$^{\cite{akkaya2020ferret}}$

C++ libraries like ROOT provide standard header files for linking their functionality to other software. Using a basic Ferret command, such a header file is imported, providing access to ROOT:
{\color{blue}\begin{verbatim}
(native-header "ROOT.h")
\end{verbatim}}
The import statement sizes the executable binary to 500KB. Also, it already shows the ubiquitous way of starting a statement in Lisp, namely with an open bracket. The second line of our translation takes in the Lisp file \texttt{cxx.clj}, which (based on Ferret’s interop features) provides higher level access to C++ libraries:
{\color{blue}\begin{verbatim}
(require '[cxx :as ROO])
\end{verbatim}}
Our code for the interop functionality, some 400 lines of self-contained Ferret code, is available free and open source.$^{\cite{kloimw2023lisroot}}$

\section{THE FUNCTION}
\subsection{Textbook notation}
The mathematical function that is plotted in the tutorial can be represented by the usual formula for a straight line.
{\color{red}
\begin{equation}
f_{d,k}(x)=d+kx
\end{equation}}
With this definition, given that we set the parameters \(\textcolor{red}{d} = 5\) and \(\textcolor{red}{k} = 2\), at point \(\textcolor{red}{x} = 0.5\) the function \textcolor{red}{$f$} has the value 6.

There is one complication though: the function used in the tutorial does not simply take the argument \textcolor{red}{$x$} and the two parameters \textcolor{red}{$d$} and \textcolor{red}{$k$} as three numbers. Instead, the actual function takes two arguments, both being vectors. The first vector is only of one dimension and carries \textcolor{red}{$x$}, the second vector is of two dimensions carrying \textcolor{red}{$d$} and \textcolor{red}{$k$}. As a result of this complication, the textbook notation for the function results in a more complicated formula.
{\color{red}
\begin{equation}
f\colon R^1 \times R^2 \rightarrow R; 
\bigl( \left( x \right), \left( d,k \right) \bigr) \rightarrow d + k x
\end{equation}}
Following the example above, if called with vectors $(0.5)$ and $(5, 2)$, the function \textcolor{red}{$f$} gives the value 6. As shown below, this formula for the function \textcolor{red}{$f$} can readily be taken over into LISP notation.

\subsection{Python Syntax}
The Python code to implement this function looks quite different from its textbook notation because, adhering to the OO programming paradigm, it involves the definition of a class, which in this case is named \texttt{\textcolor{magenta}{Linear}}:
{\color{magenta}\begin{verbatim}
class Linear:
    def __call__(self, arr, par):
        return par[0] + arr[0]*par[1]

l = Linear()
\end{verbatim}}
In OO programming, one usually does not work with classes directly, but with instances of classes. Following this pattern, an instance of the class \texttt{\textcolor{magenta}{Linear}} is created, the object named \texttt{\textcolor{magenta}{l}}. This instance, through the reserved name \texttt{\textcolor{magenta}{\_\_call\_\_}} in its class definition, can act as a function. Thus the object \texttt{\textcolor{magenta}{l}} effectively represents our function for a straight line.

\subsection{Lisp Syntax}
The Lisp code to implement the function representing a straight line looks comparatively similar to its textbook notation:
{\color{blue}\begin{verbatim}
(defn Linear []
  (fn [[x] [d k]]
    (+ d (* x k))))

(def l (Linear))
\end{verbatim}}
The most bewildering element in any Lisp code is probably the prefix notation which carries the math symbols in the “wrong” place. Specifically, we need to write \texttt{\textcolor{blue}{(+ d (* x k))}} instead of the usual \texttt{\textcolor{blue}{d + k * x}} .

For further acquaintance with this unfamiliar syntax, we give two additional examples. First, to define an unnamed function taking as arguments two numbers and returning their sum, we write \texttt{\textcolor{blue}{(fn [a b] (+ a b))}} . Second, to define a named function that takes no arguments and returns the number \texttt{\textcolor{blue}{42}}, we write \texttt{\textcolor{blue}{(defn name [] 42)}}.

In our translated Lisp code, \texttt{\textcolor{blue}{Linear}} is such a named function that takes no arguments. It is a “higher order function” in that it does not return a number, but an unnamed function (a “Lambda” in the terminology of functional programming). The syntax for calling \texttt{\textcolor{blue}{Linear}} is not \texttt{\textcolor{magenta}{Linear()}} but \texttt{\textcolor{blue}{(Linear)}}, keeping the Lisp tradition of starting with an open bracket. By calling \texttt{\textcolor{blue}{Linear}} in this way, we bind the resulting unnamed function to the name \texttt{\textcolor{blue}{l}}. So again like in the Python code, \texttt{\textcolor{blue}{l}} is our desired function representing a straight line. 

The unnamed function which \texttt{\textcolor{blue}{Linear}} returns has the signature \texttt{\textcolor{blue}{[[x] [d k]]}}. This means that this function expects two arguments A and B, both being vectors, with A = [x] and B = [d k] denoting that the elements \texttt{\textcolor{blue}{x}} and \texttt{\textcolor{blue}{d,k}} are extracted out of the vectors already within the signature.

\subsection{Industry going functional}
The translation from Python to Lisp opens a new perspective on the CERN tutorial: the given Python example can be seen as functional in disguise because the class in the tutorial effectively plays the role of a higher order function. We think that palpable function definitions are well suited for physicists and computational scientists while sophisticated concepts like the definition of classes should be left to  professional software engineers and computer scientists (note the difference between comput-ational vs. comput-er).

The replacement of the class-concept by pure functions is taking place in the wider software engineering domain. As an example we state React, the de-facto standard framework for Web development maintained by Facebook.
At React’s introduction in 2012, a Web component had to be defined in terms of a class:
{\color{magenta}\begin{verbatim}
class HelloC extends React.Component {
 render() {
  return <h1>Hi, {this.props.name}</h1>;
 }
}
\end{verbatim}}
Since the 2019 version the above OO version is only maintained for backward compatibility. In the beginner tutorial it is recommended to define the same component via the following simple function:$^{\cite{meta2023react}}$
{\color{magenta}\begin{verbatim}
function HelloF(props) {
 return <h1>Hi, {props.name}</h1>;
}
\end{verbatim}}
Indeed, also our ROOT class, this function-in-disguise, can be more cleanly written as a pure function:
{\color{magenta}\begin{verbatim}
def Linear():
    return lambda arr, par: par[0] + arr[0]*par[1]
\end{verbatim}}
In comparison, we think that the Lisp notation still better resembles the mathematical notation. However that may be, we now go on to further develop our functional scripting interface which is not interpreted but a compiled C++ layer.

\section{CALLING ROOT}
Before generating plots, an instance of ROOT’s \texttt{\textcolor{magenta}{TCanvas}}  class is created to register a canvas object in the internal state of the ROOT system.
{\color{magenta}\begin{verbatim}
c = ROOT.TCanvas()
\end{verbatim}}
The observant reader may have noticed that this creation of the object named \texttt{\textcolor{magenta}{c}} appears earlier than in the original Python code. The reason for this will be discussed later.

\subsection{Translation} 
The Lisp code for creating the canvas looks similar to the Python code:
{\color{blue}\begin{verbatim}
(def c (ROO/T new TCanvas))
\end{verbatim}}
Like in Python, also a variable named \texttt{\textcolor{blue}{c}} is created. Technically it is not an object (Ferret does not have those) but a two valued list containing the string \textcolor{blue}{”TCanvas”}  and a C++ pointer.

There is one additional element: the reserved word \texttt{\textcolor{blue}{new}}. It is required because Lisp is converted to C++ using one main interop macro, \texttt{\textcolor{blue}{ROO/T}}. Being a Lisp macro and not a function, during the conversion process the name \texttt{\textcolor{blue}{ROO/T}} completely disappears and is replaced by the generated C++ code. In contrast, \texttt{\textcolor{blue}{new}} (and of course ROOT’s \texttt{\textcolor{blue}{TCanvas}}) does not disappear but transfers over to C++ code.

\section{PLOTTING THE FUNCTION}
We stated the textbook notation of our mathematical function using the letter \textcolor{red}{$f$} while in Python the letter \texttt{\textcolor{magenta}{l}} is used for the function's name. The reason is that the colloquial name \texttt{\textcolor{magenta}{f}} has been reserved for the final Python object, an instance of a certain ROOT class called \texttt{\textcolor{magenta}{TF1}}:
{\color{magenta}\begin{verbatim}
f = ROOT.TF1('pyf2', l, -1., 1., 2)
f.SetParameters(5., 2.)
f.Draw()
\end{verbatim}}
The creation process for the ROOT object \texttt{\textcolor{magenta}{f}} can be pictured as follows: our formula \texttt{\textcolor{magenta}{l}} from before is
swallowed up and digested into an object \texttt{\textcolor{magenta}{f}} that finally represents the straight line eventually to be plotted. However, before being plotted, the object \texttt{\textcolor{magenta}{f}} is mutated using \texttt{\textcolor{magenta}{SetParameters}}: as a kind of an afterthought: we set the parameters \texttt{\textcolor{magenta}{d}} and \texttt{\textcolor{magenta}{k}} of the already digested formula to the numbers 5 and 2 respectively.

In the final statement of the code, containing the command \texttt{\textcolor{magenta}{Draw}}, the graph is plotted. Note that all three statements for plotting start with the one letter variable \texttt{\textcolor{magenta}{f}}. Starting with a variable is idiomatic in the OO paradigm which consequently insists that \texttt{\textcolor{magenta}{Draw}} is not to be termed a function but a “method of the class TF1”.

\subsection{Translation}
A one-to-one translation of the three Python plot statements to Lisp yields ungainly code with nested expressions:
{\color{blue}\begin{verbatim}
(def f ((ROO/T new TF1) "pyf2" l -1. 1. 2))
((ROO/T SetParameters TF1) f 5. 2.)
((ROO/T Draw TF1) f)
\end{verbatim}}
We can improve the notation of this nested structure by binding names to those inner expressions that start with the \texttt{\textcolor{blue}{ROO/T}} macro. This is possible because the macro is constructed such that all expressions starting with \texttt{\textcolor{blue}{ROO/T}} result in an unnamed function; so the names to be bound are going to represent functions.

While \texttt{\textcolor{blue}{ROO/T}} results in a function, everything that is within a \texttt{\textcolor{blue}{ROO/T}} expression is just a string that is handed down to C++; so is the string \texttt{\textcolor{blue}{Draw}}. But the result of a \texttt{\textcolor{blue}{ROO/T}} expression, a function, can be bound to some variable of arbitrary name and thus just as well to the name \texttt{\textcolor{blue}{Draw}}:
{\color{blue}\begin{verbatim}
(def Draw (ROO/T Draw TF1))
\end{verbatim}}
With this binding, \texttt{\textcolor{blue}{Draw}} becomes a valid Lisp function (we will see, later on, a fruitful extension of this simple binding). By adding in the same manner two respective bindings for \texttt{\textcolor{blue}{(ROO/T new TF1)}} and \texttt{\textcolor{blue}{(ROO/T SetParameters TF1)}}, and thus doubling the number of statements from three to six, we achieve the desired improved notation for plotting the graph:
{\color{blue}\begin{verbatim}
(def f (newTF1 "pyf2" l -1. 1. 2))
(SetParameters f 5. 2.)
(Draw f)
\end{verbatim}}
The doubling of statements induced by the three bindings seems worse than is actually the case. This is because all the introduced bindings, just like the one for \texttt{\textcolor{blue}{Draw}}, are completely generic, i.e. they are independent of the particular plotting problem at hand. Thus they could be isolated from the user’s code by separating them into an extra Lisp file. Taking this generic property of the bindings further, such an extra source file can be written up front, ideally by a software engineer who is familiar with the library architecture but not necessarily acquainted with every user’s specific problem domain.

\section{TOP LEVEL EXPRESSIONS}
We move on to provide a further improved notation for the above three plot statements by combining them into one single expression:
{\color{blue} \begin{verbatim}
(doto (newTF1 "pyf2" l -1. 1. 2)
  (SetParameters 5. 2.)
  Draw)
\end{verbatim}}
To perform the combination we used the standard Clojure command \texttt{\textcolor{blue}{doto}}. The three-to-one combination was only possible because all of the three plot statements pertained to the variable \texttt{\textcolor{blue}{f}}. This explains the reason for the above decision to create the canvas object early: to establish the necessary continuity for idiomatic statement- combination.

The combination just demonstrated represents the typical workflow in Lisp programming: at first lots of statements are created; they usually look ungainly but constitute working code. In a next step, facilitated by the fact that there are only functions on data, sub-expressions are extracted and statements are regrouped. Some of the resulting expressions are moved to separate files, others may span several lines (because of this multi-line spanning, Lisp code is not said to consist of statements but of “top level expressions”).

Admittedly, the disadvantage of this combining process is that computer memory is in general not used as efficiently as possible because one loses the fine-tuning ability that comes with spreading individual statements around different code blocks. The advantage is that by merely following idiomatic coding guidelines, the code for one and the same logical idea tends to gather in one place. We think that the trade-off in favor of concise notation is justified as any heavy lifting code is supposed to reside in the C++ library and thus memory consumption is of secondary consideration in LISP code.

\section{RECAP THUS FAR}
By now, we have covered the first two parts of the article. Firstly, we took a publicly available Python tutorial and completed its translation to Lisp. We argued that scientific ideas are much more directly and meaningfully expressed in Lisp’s functional programming paradigm. Secondly, we explained our Lisp-C++ interoperation method.

In the following third part of the article, we are going to further develop the interoperation method with the aim of accessing more features of ROOT’s C++ library function \texttt{\textcolor{blue}{Draw}}. This means that we are addressing the issue of interoperation with C++ “runtime polymorphism”.

\section{TYPE HINTS}
The Lisp function which is bound to the name \texttt{\textcolor{blue}{Draw}} takes only one single argument, the variable \texttt{\textcolor{blue}{f}}. But ROOT’s C++ library function \texttt{\textcolor{blue}{Draw}}, supporting e.g. plotting of dotted graphs, can take additional arguments. To hand down those optional arguments to ROOT, we need to help the interop macro \texttt{\textcolor{blue}{ROO/T}} with its C++ code-generation. This help comes in the form of creating a type hint named \texttt{\textcolor{blue}{:my-hint}}; this hint stating that the Draw method of class \texttt{\textcolor{blue}{TF1}} has a variant that takes an additional argument of type \texttt{\textcolor{blue}{string}}:
{\color{blue}\begin{verbatim}
(ROO/Ts [:TF1 :Draw :my-hint]
        [:string])
\end{verbatim}}
By calling \texttt{\textcolor{blue}{ROO/T}} with \texttt{\textcolor{blue}{:my-hint}}, any graph can be dotted point by point, the value of the respective argument for ROOT being the letter \texttt{\textcolor{blue}{P}}:
{\color{blue}\begin{verbatim}
((ROO/T Draw TF1 :my-hint) f "P")
\end{verbatim}}
In general, the macro \texttt{\textcolor{blue}{ROO/T}} always needs a hint to generate C++ code; we just did not up to now encounter any hints because the interop system loads appropriate default settings. Those default settings can also be overridden by giving a newly created Schema the name \texttt{\textcolor{blue}{:default}} (instead of e.g. \texttt{\textcolor{blue}{:my-hint}}).

\section{RUNTIME CHECKS}
Our interop software additionally provides an important sophistication of the type-hint concept: runtime value validation. The following more sophisticated hint does not only contain the information that ROOT’s C++ function \texttt{\textcolor{blue}{Draw}} accepts an additional argument of type \texttt{\textcolor{blue}{string}}, but that at runtime we want this option to have the name \texttt{\textcolor{blue}{style}} and that its runtime value always only contains one letter:
{\color{blue}\begin{verbatim}
(ROO/Ts [:TF1 :Draw :your-hint]
        [:string]
        [[:style ::one-letter]])
\end{verbatim}}
This form of hint goes beyond the specification of mere data types facilitating the generation of C++ code. It not only contains value types for compile time but in addition specifies names and restrictions for value ranges to be obeyed at runtime, thus deserving the label “Data Schema”.

To be sure, for the time being, when plotting a dotted graph using either \texttt{\textcolor{blue}{:my-hint}} or \texttt{\textcolor{blue}{:your-hint}} makes little difference:
{\color{blue}\begin{verbatim}
((ROO/T Draw TF1 :my-hint) f "P")
((ROO/T Draw TF1 :your-hint) f {:style "P"})
\end{verbatim}}
But the specification of the runtime checked data Schema \texttt{\textcolor{blue}{:your-hint}} has the following far reaching consequence: if the data at runtime does not match the Schema, instead of calling ROOT, a mismatch code is returned. This automatic error handling allows the fruitful extension of a simple binding already shown above, which was
{\color{blue}\begin{verbatim}
(def Draw (ROO/T Draw TF1))
\end{verbatim}}
The extension of this already known \texttt{\textcolor{blue}{Draw}} is the following new function \texttt{\textcolor{blue}{fallbackDraw}} which, in case of a runtime mismatch, resorts to a fallback:
{\color{blue}\begin{verbatim}
(defn fallbackDraw [f params]
  (when (:mismatch
         ((ROO/T Draw TF1 :your-hint) f
          params))
    ((ROO/T Draw TF1) f)))
\end{verbatim}}
To plot the dotted graph, we do as expected: we call the just created fallback-enabled function by its name \texttt{\textcolor{blue}{fallbackDraw}}, resulting in the execution of the ROOT C++ code drawing a dotted graph:
{\color{blue}\begin{verbatim}
(fallbackDraw f {:style "P"})
\end{verbatim}}
Now, when we draw not a dotted but a continuous graph by again calling the function \texttt{\textcolor{blue}{fallbackDraw}} but with different arguments,
{\color{blue}\begin{verbatim}
(fallbackDraw f {:style "unknown"})
\end{verbatim}}
a different portion of compiled code is executed. This is because the string \texttt{\textcolor{blue}{"unknown"}} is more than one letter, thus failing the runtime check and leading to the execution of the fallback which draws the ROOT default of a continuous graph.

So, while in all cases before, calling a specific function always resulted in the execution of the same C++ code, now calling the same name twice can result in the execution of different parts of pre-compiled C++ code.

At first glance, the fallback function just presented looks very innocent, but in fact it represents the blueprint for a Lisp “multimethod” function that is able to provide access to all of ROOT’s \texttt{\textcolor{blue}{Draw}} capabilities without the need of user provided Schema hints. Instead of relying on user information, the function checks the users’ input data before dispatching to the according C++ code, the already mentioned mechanism called “runtime polymorphism”. To fully grasp its blueprint quality, it is important to realize that the number of Schemas and checks within a function, like that of optional arguments, is not fixed but the choice of a software engineer who ideally shares a set of generic multimethods with other users working in their specific domains.

\section{SCHEMA IN A FILE}
We just demonstrated how to declare a specific Schema within the user’s code, a method akin to the way access to external libraries is provided by the Python-to-C compiler Cython.$^{\cite{behnel2010cython}}$ But while in Cython the code for declaring an external function resembles that of a normal function declaration, within \texttt{\textcolor{blue}{ROO/T}}, such a declaration is made via a data structure. More specifically, the Schemas \texttt{\textcolor{blue}{:my-hint}} and \texttt{\textcolor{blue}{:your-hint}} above, with their idiomatic abundance of colons, are representations of a Clojure- standard for data Schema definitions: Malli data structures.$^{\cite{metosin2023malli}}$ Insofar as they resemble XML data structures, their combination with multimethods has similarities with the XML/function-stub strategy used by ROOT before the introduction of cppyy\footnote{"Instead of optimizing existing concepts we decided to develop a new C++ interpreter"$^{\cite{naumann2008role}}$}.

The Malli data structure for a Schema is used twice: once during compile time to generate C++ code and once during runtime to validate the data and possibly dispatch to different parts of the pre-compiled code (runtime polymorphism). Representing the type information as a data structure guarantees the necessary easy access to the Schemas during runtime.

This “Schema as Data” approach naturally lends itself to storing Schema declarations in an external text file. Indeed, the default Schemas for the code-examples of this article are stored in a separate machine- as well as human-readable text file called \texttt{malli\_types.edn} . It does not only contain Schemas for C++ classes and methods, but, to facilitate the respective callback form C++, also a type signature for the Lisp function \texttt{\textcolor{blue}{Linear}}.

\section{SCOPE AND LIMITATIONS}
The interop system presented here is not only aimed at Cern’s ROOT library, but is a generic tool intended to access arbitrary C++ libraries. To make this point, the information in \texttt{malli\_types.edn} is sufficient to create and manipulate a text object of the standard C++ \texttt{std::string} class using the \texttt{\textcolor{blue}{ROO/T}} macro (which consequently is a misleading name in this case). However, the interop is limited to creating objects (not garbage collected) and calling their methods which in their turn can only return primitive types. Not only is the marshalling feature- incomplete, also its speed is not optimized: preliminary benchmarks for a numerical root search of Lisp functions show, compared to native C++ functions, a performance penalty of two orders of magnitude. Also only a handful of Schemas are provided and even such a basic functionality like accessing object properties is lacking.

We used Ferret as our Lisp-compiler because jank was not ready at the time of writing. Moreover we hope that a future Lisp-to-C++ compiler will, inspired by examples from the games industry,$^{\cite{liebgold2011functional}}$ be able to compile to native C++ types so that the performance penalty of marshalling will disappear. As such, this article is for demonstrating this one main point: a binary compiled functional layer for C++ allowing concise Lisp notation for runtime polymorphism. Nevertheless, if maintainers of a C++ library wrote a Malli structure with Schema hints for its classes and methods, then those could readily be accessed with the presented interop system. In general, the hints that refer to the respective pre-defined Schemas need to be provided within the user code. But as shown in the fallback example, if the maintainers also write additional Lisp multimethod functions, a user wanting to access some particular library function will not have to specify any type hints at all.

\section{CONCLUSION AND OUTLOOK}
We used a simple example from CERN's ROOT library tutorial to directly compare Lisp Code based on the functional programming paradigm with Python code following the object-oriented approach. The example clearly shows that the functional code is more intuitive and expressive for scientists. Next, we showed how to put this claim into practice by developing a way to access ROOT via the Lisp dialect Clojure. For this, using the Ferret compiler, Lisp was used to  compile a C++ layer providing a functional interface to ROOT. The type hints necessary for C++ code generation were stored in a separate data file. For seamless interoperation without obstructing the Lisp syntax, the use of statically compiled “multimethods” was demonstrated. Those perform runtime checks on the input data and automatically dispatch to the respective C++ functionality (runtime polymorphism). The type hints are stored as a Malli data structure to ensure compatibility with the larger Clojure ecosystem.

The interop system presented here is a generic framework that, when provided with suitable type hints, facilitates access to methods of arbitrary C++ libraries, which are widely used in science, engineering, and quantitative finance. All necessary hint-information about classes and methods need to be written into a machine- readable Malli text file. It is conceivable to in the future create the necessary type hints and multimethods automatically by tapping into the yet unused "automatic binding" features of ROOT's cppyy infrastructure. Also, it is conceivable to code and statically compile a kind of multimethod that is suited to be directly called from languages like e.g. Julia$^{\cite{eschle2023potential}}$.

In a broader context, we showed how to glue Lisp onto a C++ library. We agree with YCombinator's founder Graham: "You can write little glue programs in Lisp too ..., but the biggest win for languages like Lisp is at the other end of the spectrum, where you need to write sophisticated programs to solve hard problems...".$^{\cite{graham2002revenge}}$

Every Physicist has to do plumbing work but eventually will encounter some  hard problem. Using Lisp from the start as glue language makes sure to have this useful tool available when, as described in “The role of programming in the formulation of ideas”$^{\cite{sussman2002role}}$, hard problems arrive.

\section{ACKNOWLEDGMENTS}
We thank J. Wilkerson for fruitful discussions.

\bibliographystyle{unsrt}

\bibliography{refs}

\end{document}